# Study of cosmic ray sources using data on extragalactic diffuse gamma-ray emission

A. Uryson

Lebedev Physical Institute RAS, Moscow

E-mail: uryson@sci.lebedev.ru

**Abstract**. We discuss ultra-high energy (UHE) cosmic rays (CR) from minor sources and their possible contribution to the extragalactic diffuse gamma-ray emission. As an illustration of minor sources we consider possible specific type of active galactic nuclei (AGN) in which supermassive black hole is surrounded by a super strong magnetic field of $10^{10} - 10^{11}$ Gs. In this model we have calculated CR energy spectra at the Earth and intensity of cascade quanta produced by CRs in extragalactic space. Proceeding from numerical results and the data by Pierre Auger Observatory and Telescope Array it is shown that these AGNs make a negligible contribution to the UHECR flux at the Earth. However CRs from the AGNs produce significant gamma-ray flux as compared to extragalactic diffuse gamma-ray emission measured by Fermi LAT. We conclude that UHECRs from minor sources can contribute noticeably to extragalactic diffuse gamma-ray emission.

## 1. Introduction

Sources of ultra-high energy (UHE) cosmic rays (CRs) seems to be extragalactic. Evidently they are active galactic nuclei (AGN) but they are not still revealed. Source identification by particle arrival directions was not successful. A number of reasons makes identification difficult. First, identification is carried out assuming that in extragalactic space particles propagate almost rectilinearly. However UHECRs are presumably deflected in magnetic fields. Next, error-boxes of ~$1^0$ around arrival directions result in a large number of astrophysical objects in area near the particle arrival and this impedes source identification.

In extragalactic space, UHECRs lose energy while interacting with the cosmic microwave background. This results in two effects: a suppression of the UHECR energy spectrum if UHECRs come from distances of more than approximately 50 Mpc - GZK-effect [1, 2] and in electromagnetic cascades in extragalactic space [3, 4]. Currently both UHECR spectra and cascade gamma-ray emission are studied to investigate UHECR sources. Data on spectra are obtained with CR giant arrays - Pierre Auger Observatory (PAO) and Telescope Array (TA). To analyze cascade emission Fermi LAT data [5] are used.

The common way to examine UHECR sources is to describe the particle energy spectrum measured and next to analyze emission produced by particles in extragalactic space. The intensity of cascade gamma-quanta is required to be less than the measured intensity of extragalactic diffuse gamma-ray emission (excluding contribution of unresolved gamma-ray sources). Models with parameters matching this requirement are selected for further analyze (see e.g. [6-8]).

In this work we consider possible specific type of AGNs in which supermassive black hole is surrounded by a super strong magnetic field of $10^{10} - 10^{11}$ Gs. Particles are accelerated there to ultra-high energies [9-12]. Here we show that these AGNs contribute negligibly to the UHECR flux at the Earth. However UHECRs generate noticeable diffuse gamma-ray flux that amounts to few tens percent of the flux measured by Fermi LAT (excluding contribution of unresolved gamma-ray sources). The same result was obtained previously in [13] using semi quantitative estimate. We conclude that UHECRs from minor sources can contribute noticeably to extragalactic diffuse gamma-ray emission. This is important when studying both UHECR source models and dark matter decays that add to diffuse gamma-ray emission.

Computing was performed with the code TransportCR [14].

## 2. The model

Model assumptions concern following points: injection spectra and cosmic evolution of CR sources, extragalactic background emission, and the extragalactic magnetic field.

We assume that CR sources are point. Next we assume that UHECR sources are AGNs with super strong magnetic fields where protons are accelerated in induced electric fields to energies of $10^{21}$ eV. Due to acceleration mechanism it is reasonable to assume that the initial CR spectrum is monoenergetic with the energy $E=10^{21}$ eV.

We assume also that CRs consist of protons.

Cosmic evolution of objects discussed is not clear. We analyze two possible cases of source evolution: as evolution of Blue Lacertae objects (Bl Lac's) [14, 15] and as evolution of radio AGNs described in [16]. In the latter case only source density evolution is considered here (as it is not clear how luminosity of objects with super strong magnetic field varies with redshift).

In the model, background emission is treated in the following way.

Microwave background emission has Planck distribution in energy with the mean value $\varepsilon_r=6.7\cdot10^{-4}$ eV, the mean photon density is $n_r=400$ cm$^{-3}$.

Extragalactic background light has characteristics described in [17].

For the radio background emission the model [18] with the pure luminosity evolution for radio galaxies is used.

Extragalactic magnetic fields are weak enough so cascade electrons loose negligible energy in synchrotron radiation ([14] and references therein).

## 3. Results

Calculated UHECR spectra along with the spectrum obtained at PAO are shown in Fig. 1. Calculated spectra are normalized to the PAO spectrum at the energy of $10^{19.5}$ eV.

In both cases of source evolution model spectra differ significantly in shape from spectra measured at PAO (as well as TA). Spectra calculated are several orders lower.

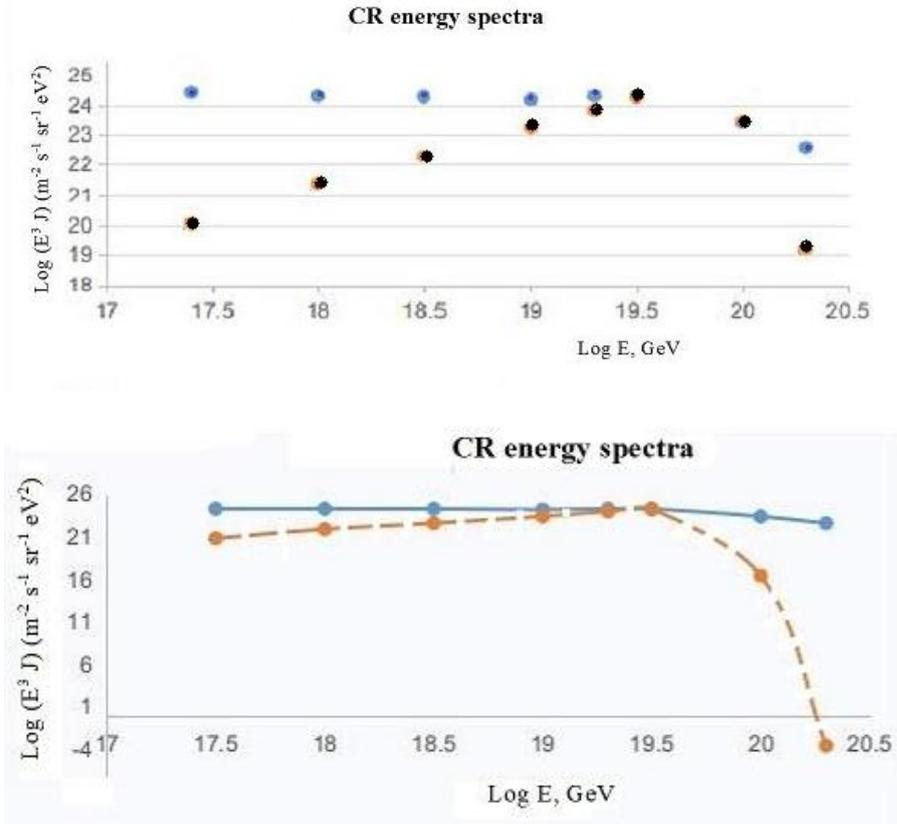

Fig. 1: CR energy spectra. *Top panel:* fit of PAO CR spectrum [19] (blue points) and the calculated UHECR spectrum with source *z*-dependence as of BL Lac's [14, 15]. *Bottom panel:* fit of PAO CR spectrum [19] (blue solid line) and the calculated UHECR spectrum with source *z*-dependence for pure density evolution [16] (red dashed line).

We proceed now to the integral intensity of gamma rays above 50 GeV produced by UHECRs in extragalactic space. We choose this energy range as contribution of unresolved gamma-ray sources is obtained for it [20], and the contribution is taken into account when comparing with Fermi LAT data.

The integral intensity of cascade gamma rays is:

$I\gamma$ ($E$>50 GeV, source z-dependence: BL Lac's [14, 15]) = $1.002 \times 10^{-9}$ (cm$^{-2}$ s$^{-1}$ sr$^{-1}$),  (3.1)

$I\gamma$ ($E$>50 GeV, source z-dependence: [16]) = $1.641 \times 10^{-9}$ (cm$^{-2}$ s$^{-1}$ sr$^{-1}$).  (3.2)

The difference in values results from dissimilar z-dependences used in calculation.

## 4. Discussion

Now we compare the integral intensity obtained with Fermi LAT data [5]. Truly diffuse extragalactic gamma-ray background - IGRB (the isotropic diffuse gamma-ray background) is produced by electromagnetic cascades, which high-energy gamma rays and UHECRs initiate in extragalactic space. IGRB obtained by Fermi LAT includes emission from unresolved individual extragalactic sources. At energies above 50 GeV, their contribution equals to 86 percent [20].

From [5]
$$\text{IGRB }(E>50\text{ GeV}) = 1.325 \times 10^{-9} \text{ (cm}^{-2}\text{ s}^{-1}\text{ sr}^{-1}) \quad (4.1)$$
and excluding contribution of unresolved sources we obtain
$$\text{IGRB }_{\text{without blazars}}(E>50\text{ GeV}) = 1.85 \times 10^{-10} \text{ (cm}^{-2}\text{ s}^{-1}\text{ sr}^{-1}). \quad (4.2)$$
This value is much less than those calculated in the model.

Accounting for measurement errors, uncertainties in the model of the diffuse galactic emission (which is used to obtain IGRB from Fermi LAT data), and uncertainty in the value of unresolved source contribution (the contribution can be 12 percent lesser) the value of IGRB$_{\text{without blazars}}$ ($E>50$ GeV) lies in the band:
$$2.20 \times 10^{-10} \text{ (cm}^{-2}\text{ s}^{-1}\text{ sr}^{-1}) \leq \text{IGRB }_{\text{without blazars}}(E>50\text{ GeV}) \leq 5.40 \times 10^{-10} \text{ (cm}^{-2}\text{ s}^{-1}\text{ sr}^{-1}). \quad (4.3)$$
The values calculated in the model are several times higher as before.

In the model integral intensity of cascade gamma rays is derived normalizing UHECR spectra at the energy of $10^{19.5}$ eV. However minor sources give negligible contribution to the UHECR flux. Thus it is more natural to normalize spectra calculated to another value. As an example way of example we choose *ad hoc* the normalizing value equals about 20 percent of the previous normalizing value. Then the integral intensity of cascade gamma rays equals:
$$I\gamma\ (E>50\text{ GeV, source z-dependence: BL Lac's [14, 15]}) \approx 2.0 \times 10^{-10} \text{ (cm}^{-2}\text{ s}^{-1}\text{ sr}^{-1}), \quad (4.4)$$
$$I\gamma\ (E>50\text{ GeV, source z-dependence: [16]}) \approx 3.3 \times 10^{-10} \text{ (cm}^{-2}\text{ s}^{-1}\text{ sr}^{-1}). \quad (4.5)$$
These values fall within the interval of measured IGRB (4.3), leaving room for contribution of UHECRs from dominant sources to the diffuse gamma-ray background.

## 5. Conclusion

We consider UHECRs from possible specific type of AGNs in which supermassive black hole is surrounded by a super strong magnetic field of $10^{10} - 10^{11}$ Gs. In these sources protons are accelerated in induced electric fields to the energy of $10^{21}$ eV [9, 10, 12]. Due to mechanism of acceleration it is reasonable to assume that the initial CR spectrum is monoenergetic with the energy $E=10^{21}$ eV.

Cosmic evolution of AGNs discussed is not clear. We analyze two possible cases of AGN evolution: as evolution of Bl Lac's [14, 15] and as evolution of radio AGNs described in [16].Computing was performed with the code TransportCR [14].

UHECR spectra calculated are several orders lower than that measured at PAO and TA. However CRs from the AGNs discussed appear to produce significant gamma-ray flux - of several tens percent of IGRB measured by Fermi LAT [5].

AGNs discussed illustrate possible role of UHECRs from minor sources in contributing to diffuse gamma-ray emission. We conclude that UHECRs from minor sources can give noticeable of extragalactic diffuse gamma-ray background.

**Acknowledgment**
The author is grateful to O. Kalashev for discussion.